# RESEARCH ARTICLE

## Non-singular arbitrary cloaks dressing three-dimensional anisotropic obstacles

G. Dupont, S. Guenneau[*] and S. Enoch

*Institut Fresnel, UMR CNRS 6133, University of Aix-Marseille,
case 162, F13397 Marseille Cedex 20, France*

We design three dimensional electromagnetic cloaks, starting from a small region of complex shape instead of a point. We derive the expression of a transformation matrix describing an objet with a surface of revolution and its associated non-singular cloak. We note that while none of the eigenvalues vanish inside the cloak, they suffer a discontinuity on its inner surface. Moreover, all three eigenvalues are independent upon the radius in the concealed object. The validity of our analytical results is confirmed by finite edge-elements computations showing scattering is much reduced when the object is dressed with the cloak. We note that neither the object nor the cloak are invisible on their own.

## 1. Introduction

Structured metamaterials based upon resonant structural elements (*1*) have gained popularity over the past few years due to their counter-intuitive physics as well as their scope for subwavelength imaging applications (through negative refractive effective index (*2*)) and invisibility cloaks (via artificial magnetism and anisotropic heterogeneous features). Transformational optics is a vibrant area in metamaterial research based on newly engineered heterogeneous anisotropic media (*3, 4*), low index materials (*5*) and of course negative refractive index materials leading to cloaking via anomalous resonances (*6*). Interestingly, the former type of invisibility is preserved in the case of an extreme near field (*7*), when the ray optics picture breaks down. Importantly, the mathematical foundations behind the scene have been known from researchers working in the area of inverse conductivity problems (*8*). The first experimental realization of an invisibility cloak, chiefly achieved in the microwave regime (*1*), suggests that cloaking will be limited to a very narrow range of frequencies. However, it will not be perfect since the cloak is necessarily dissipative and dispersive, and some of its tensor components are singular on the inner boundary. This drawback can be overcome by considering a generalized transform (*9, 10*) in an upper dimensional space and then projecting the resulting metric on the physical space, and this leads to non-singular tensors of permittivity and permeability. Alternatively, one can consider the geometric transform of Kohn, Shen, Vogelius and Weinstein (*11*) which maps a ball of radius $r_0$, $B(0, r_0) = \{r \geq 0 \mid r < r_0\}$, onto a hollow ball $B(0, R_2) - \overline{B(0, R_1)} = \{r \geq 0 \mid R_1 < r < R_2\}$. The singularity in the parameters of the cloak is controlled by $r_0$: the smaller $r_0$, the larger the interval of electromagnetic parameters. In the limit when $r_0$ vanishes, the transform amounts to mapping a point onto the hollow ball $B(0, R_2) - \overline{B(0, R_1)}$, as in (*3*), and the cloak becomes singular. The price to pay in the approach by Kohn et al. is that we now have to consider a heterogeneous anisotropic object filling the overall ball $B(0, R_1)$, whereas in the approach by Pendry et al. (*3*), this ball can be filled with any kind of object (but the cloak is singular on its inner boundary $r = R_1$). It seems fair to say that the latter approach has been extensively studied in the literature. Alternatively, one can design an approximate structured cloak via homogenization (*12–14*). The rapid growth of the field, fueled by a keen interest of the optics community, promises a large panel of new technological applications.

---

[*]Corresponding author. Email: sebastien.guenneau@fresnel.fr

A natural question to ask is whether one can design cloaks of non-spherical shapes, which are non-singular, generalising the transform by Kohn et al. (*11*). In the present paper, we discuss such a design of three-dimensional cloaks with an arbitrary cross-section described by two functions $R_1(\phi)$ and $R_2(\phi)$ giving an angle dependent distance from the origin. These functions correspond respectively to the interior and exterior boundary of the cloak. We shall only assume that these two boundaries can be represented by a differentiable function. Their finite Fourier expansions are thought in the form $R_j(\phi) = a_{0,0}^j + \sum_{n=1}^{p} a_{0,n}^j \cos(n\phi)$, $j = 1, 2$, where $p$ can be a small integer, and $a_{0,n}^j = 0$ for $n \neq p$ for computational easiness. To illustrate our methodology, we compute the electromagnetic field diffracted by a cloak with rotational symmetry about the $z$-axis. We perform full-wave finite element simulations in the commercial package COMSOL, when the cloak is illuminated by an approximate plane wave (generated by a constant electric field on the upper surface of the computational domain).

## 2. Cloaking from a nutshell

The touchstone of this paper is the blowup of a small region (a nutshell), and not a point, to create a hole in the transformed electromagnetic space. This means an object surrounded by the resulting cloak should scatter an electromagnetic wave in the same way as the nutshell (provided specific boundary conditions hold on the surface of the nutshell e.g. Neumann boundary conditions).

The geometric transformation which maps the field within the domains $\rho \leq R_1(\phi)$ and $R_1(\phi) \leq \rho' \leq R_2(\phi)$ onto itself can be expressed as:

$$\rho'(\rho, \phi) = \begin{cases} \dfrac{R_1(\phi)}{r_0}\rho & (\rho \leq R_1(\phi)) \\ \dfrac{R_2(\phi) - R_1(\phi)}{R_2(\phi) - r_0}\rho + R_2(\phi) \cdot \dfrac{R_1(\phi) - r_0}{R_2(\phi) - r_0} & (R_1(\phi) \leq \rho \leq R_2(\phi)) \end{cases} \quad (1)$$

with $\theta' = \theta$, $0 < \theta \leq 2\pi$, $\phi' = \phi$, $-\pi/2 < \phi \leq \pi$. We note that the transformation maps the field for $\rho > R_2(\phi)$ onto itself through the identity transformation. We note that $\rho'$ is a continuous piecewise smooth. Importantly, the present transform generalizes significantly the design of singular (i.e. when $r_0 = 0$) three-dimensional cloaks with a surface of revolution reported in (*15*) whereby $\rho'(\rho, \phi) = R_1(\phi) + \rho \dfrac{R_2(\phi) - R_1(\phi)}{R_2(\phi)}$ for $0 \leq \rho \leq R_2(\phi)$.

This change of co-ordinates (1) is characterized by the transformation of the differentials through the Jacobian:

$$\mathbf{J}(\rho', \phi') = \frac{\partial(\rho(\rho', \phi'), \theta, \phi)}{\partial(\rho', \theta', \phi')} \ . \qquad (2)$$

This amounts to replacing a homogeneous isotropic medium with scalar permittivity and permeability $\varepsilon$ and $\mu$, by a metamaterial described by anisotropic heterogeneous matrices of permittivity and permeability given by

$$\underline{\underline{\varepsilon}}' = \varepsilon \mathbf{T}^{-1} , \quad \text{and} \quad \underline{\underline{\mu}}' = \mu \mathbf{T}^{-1} , \qquad (3)$$

where $\mathbf{T} = \mathbf{J}^T \mathbf{J}/\det(\mathbf{J})$ is a representation of the metric tensor in the stretched radial coordinates. Importantly, there is no change in the impedance of the media since the permittivity and permeability undergo the same transformation: $Z^2 = \mu/\varepsilon = \underline{\underline{\mu}}' \underline{\underline{\varepsilon}}'^{-1}$.

After some elementary algebra, we find that

$$\mathbf{T}^{-1} = \begin{pmatrix} \frac{c_{13}^2 + \rho^2}{c_{11}\rho'^2} & 0 & -\frac{c_{13}}{\rho'} \\ 0 & c_{11} & 0 \\ -\frac{c_{13}}{\rho'} & 0 & c_{11} \end{pmatrix} \qquad (4)$$

where:
$$\begin{cases} c_{11} = \dfrac{r_0}{R_1(\phi')} \\ c_{13} = -\dfrac{r_0 \rho'}{R_1(\phi')^2} \cdot \dfrac{dR_1(\phi')}{d\phi'} \end{cases} \quad \text{when} \quad \rho \leq R_1(\phi) \text{ and}$$

$$\begin{cases} c_{11} = \dfrac{R_2(\phi') - r_0}{R_2(\phi') - R_1(\phi')} \\ c_{13} = \dfrac{(r_0 - R_2(\phi'))(R_2(\phi') - \rho')}{(R_2(\phi') - R_1(\phi'))^2} \cdot \dfrac{dR_1(\phi')}{d\phi'} + \dfrac{(r_0 - R_1(\phi'))(\rho' - R_1(\phi'))}{(R_2(\phi') - R_1(\phi'))^2} \cdot \dfrac{dR_2(\phi')}{d\phi'} \end{cases}$$

when $R_1(\phi) \leq \rho \leq R_2(\phi)$.

We note that $c_{11}$ and $c_{13}$ are discontinuous at $\rho = R_1$. This is easily seen in Fig. 1. Elsewhere, $\mathbf{T}^{-1}$ reduces to the identity matrix ($c_{11} = 1$ and $c_{13} = 0$ for $\rho' > R_2(\phi')$).

## 3. Eigenvalue analysis of the transformation matrix

To exemplify the jump in the material parameters at the interface between the concealed object and its cloak, we show the variation of $(T^{-1})_{11}$ in Fig. 1(d). We note that this coefficient varies between 0.366 and 1.35. We further observe that none of the other coefficients vanish in Fig. 1(a)-(c), hence showing that neither the cloak nor the object have singular parameters.

In the case of a spherical cloak, $c_{13}$ vanishes and $\mathbf{T}^{-1}$ reduces to $\text{Diag}(\frac{\rho^2}{c_{11}\rho'^2}, c_{11}, c_{11})$, where $c_{11} = r_0/R_1$ when $\rho \leq R_1$ and $c_{11} = (R_2 - r_0)/(R_2 - R_1)$, when $R_1 \leq \rho \leq R_2$. We note that the first eigenvalue vanishes on the inner boundary and the other two remain constant, which is consistent with the singularity analysis led in (*11*). This is unlike the circular cylindrical case whereby one eigenvalue goes to zero while the other one goes to infinity on the inner boundary (*7*).

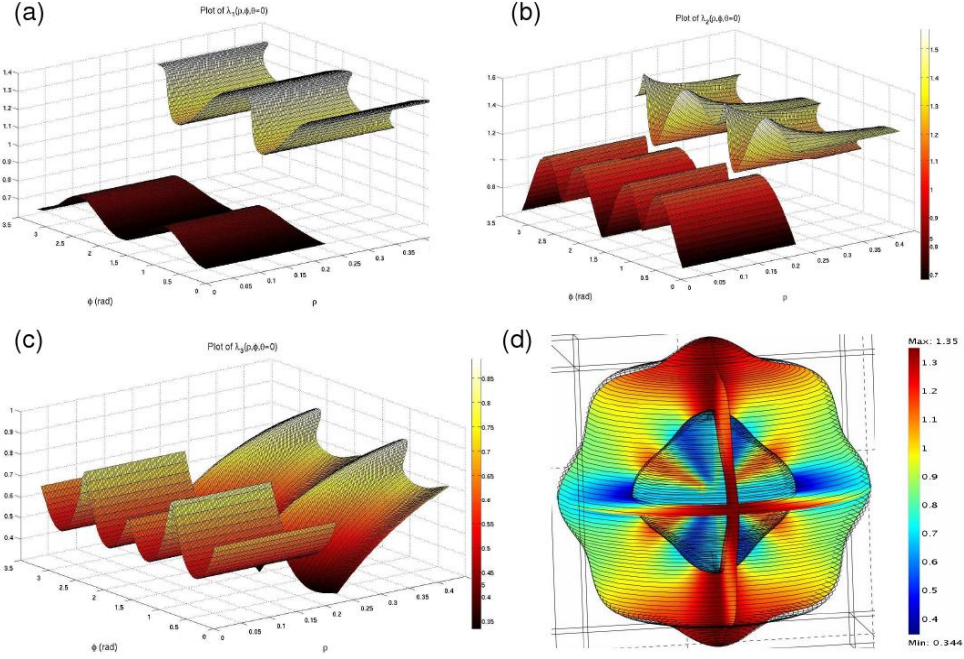

Figure 1. 3D plot of the eigenvalues of the transformation matrix: (a) 3D plot of $\lambda_1$ versus $\rho$ and $\theta$; (b) 3D plot for $\lambda_2$; (c) 3D plot for $\lambda_3$; (d) $(T^{-1})_{11}$, as given by Eq. 4, within the cloak and the object concealed inside, with boundaries given by Eq. 7.

However, in our case, the cloak is of an arbitrary shape, and it is therefore illuminating to look at the behaviour of its permittivity and permeability tensors's eigenvalues. The eigenvalues of (4) are found to be

$$\lambda_j = \frac{c_{13}^2 + \rho^2 + c_{11}^2 \rho'^2}{2c_{11}\rho'^2} + \frac{(-1)^j}{2}\sqrt{\left(\frac{c_{13}^2 + \rho^2 + c_{11}^2 \rho'^2}{c_{11}\rho'^2}\right)^2 - 4\frac{\rho^2}{\rho'^2}} \ , \ j = 1, 2 \ , \quad (5)$$
$$\lambda_3 = c_{11} \ .$$

We can therefore see that $\lambda_j$, $j = 1, 2$ are spatially varying functions of $\rho'(\rho, \phi)$, such that $\lambda_i > 0$ when $\rho = R_1$ i.e. at the inner boundary. Last, we checked that when there is no longer a symmetry of revolution about one axis, all three eigenvalues are also spatially varying with $\theta'$, but $\lambda_3$ remains independent upon $\rho$. Now, the expression of the eigenvalues further simplifies in the object:

$$\lambda_1 = \frac{r_0}{R_1(\phi)} \text{ and } \lambda_{2,3} = \frac{(\partial_{\phi'} R_1)^2 \rho^2 r_0^2 + 2\rho^2 R_1^2 r_0^2 \pm \sqrt{(\partial_{\phi'} R_1)^2 \rho^4 \left((\partial_{\phi'} R_1)^2 + 4R_1^2\right) r_0^4}}{2\rho^2 R_1^3 r_0}$$
$$= r_0 \frac{(\partial_{\phi'} R_1)^2 + 2R_1^2 \pm \sqrt{(\partial_{\phi'} R_1)^2 \left((\partial_{\phi'} R_1)^2 + 4R_1^2\right)}}{2R_1^3} \ .$$

Remarkably, all three eigenvalues go to zero linearly with $r_0$ as it vanishes, and there are all independent upon $\rho$. In the case of a spherical cloak, all three eigenvalues are constant and equal: $\lambda_j = r_0/R_1$, $j = 1, 2, 3$. The object becomes therefore ho-mogeneous isotropic (but is still magnetic). This reflects the fact that the geodesics for light within the arbitrary shaped cloak follow more and more complex trajectories when we perturb the geometry away from the spherical cloak, and need be compensated by more and more complex trajectories in the object, so that the optical system formed by the object and the cloak is invisible on the whole. Unlike for other approaches, the cloak itself is not invisible.

## 4. Finite element computations of the scattered field

We would now like to further investigate the electromagnetic response of the object and its cloak to an incident plane wave from above, see Fig. 2. For this, we choose the electric field $\mathbf{E}$ as the unknown:

$$\nabla \times \left(\underline{\underline{\mu'}}^{-1} \nabla \times \mathbf{E}\right) - k^2 \underline{\underline{\varepsilon'}} \mathbf{E} = \mathbf{0} \tag{6}$$

where $k = \omega\sqrt{\mu_0 \varepsilon_0} = \omega/c$ is the wavenumber, $c$ being the speed of light in vacuum, and $\underline{\underline{\varepsilon'}}$ and $\underline{\underline{\mu'}}$ are defined by Eqs. (3). Also, $\mathbf{E} = \mathbf{E}_i + \mathbf{E}_d$, where $\mathbf{E}_i$ is the incident field and $\overline{\mathbf{E}}_d$ is the diffracted field which satisfies the usual outgoing wave conditions (to ensure existence and uniqueness of the solution). The weak formulation associated with Eq. 6 is discretised using second order finite edge elements (or Whitney forms) which behave nicely under geometric transforms (*16*).

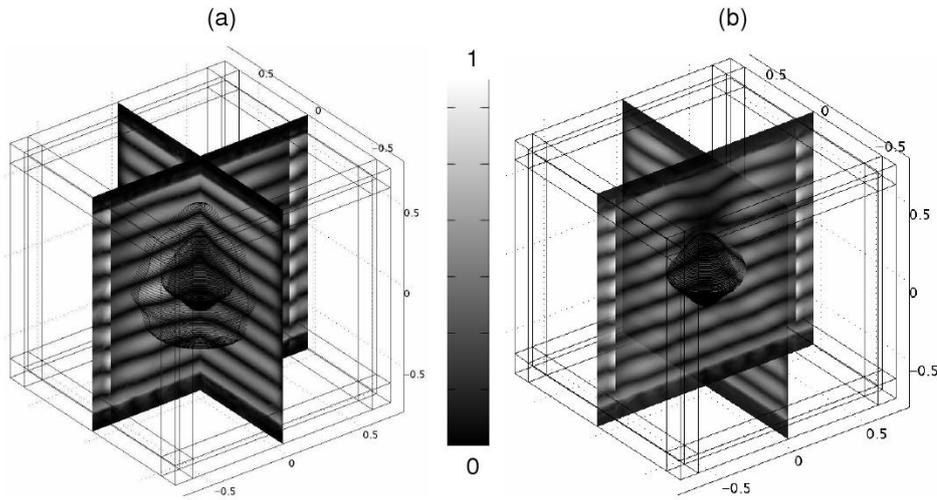

Figure 2. 3D plot of the magnitude $\sqrt{E_1^2 + E_2^2 + E_3^2}$ of the total electric field for a plane wave of wavenumber $k = 2\pi/0.3$ incident from below on a heterogeneous object with anisotropic permittivity and permeability on its own (b) and surrounded by an invisibility cloak (a).

For the sake of illustration, *cf.* Fig. 2, let us consider a cloak with inner and outer boundaries expressed as

$$\begin{aligned} R_1(\theta) &= 0.2 + 0.02\cos(4\phi) \;, \\ R_2(\theta) &= 0.4 + 0.02\cos(8\phi) \;. \end{aligned} \tag{7}$$

We report the computations for the magnitude of the total electric field in Fig. 2 (3D plot), 3 (2D plot in the $xz$-plane) and 4 (2D plot in the $yz$-plane) for a plane wave incident from above at wavenumber $k = 2\pi/0.3$ (units are in inverse of a length, say $\mu$m$^{-1}$ for nearly visible light (UV)). Around $4.10^5$ elements were used in this computation, which corresponds to about $2.7\ 10^6$ degrees of freedom. While the convergence of the numerical scheme has been checked by considering different types of meshes, the large size of the system means we were not able to further refine the mesh with the computational resources at hand.

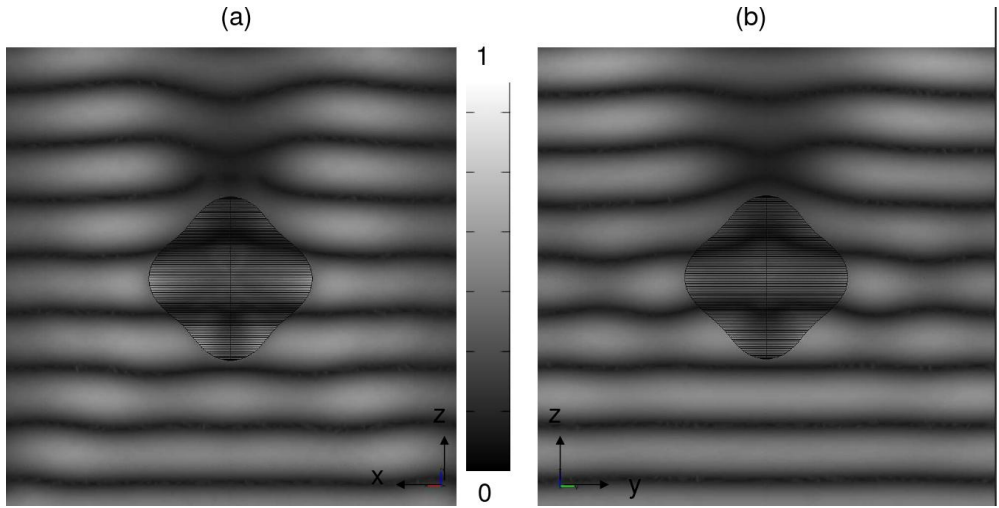

Figure 3.  2D plot of $\sqrt{E_1^2 + E_2^2 + E_3^2}$ generated by a slice of Fig. 2 in the $xz$ and $yz$-planes for $y = 0$. Plane wave incident from below on a heterogeneous object with anisotropic permittivity and permeability.

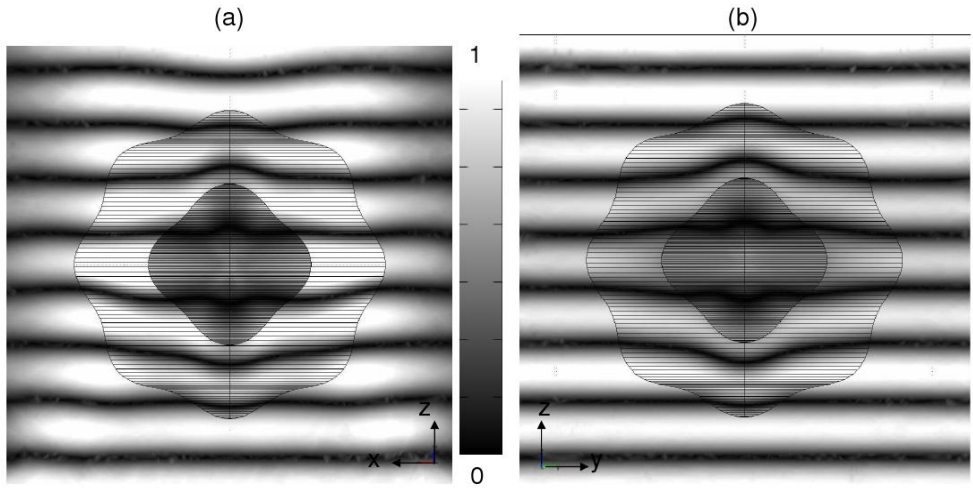

Figure 4.  2D plot of $\sqrt{E_1^2 + E_2^2 + E_3^2}$ generated by a slice of Fig. 2 in the $xz$ and $yz$-planes for $y = 0$. Plane wave incident from below on a heterogeneous object with anisotropic permittivity and permeability surrounded by the non singular cloak.

## 5.  Conclusion

In conclusion, we have proposed a design of an arbitrarily shaped non-singular cloak using a generalization of Kohn's geometric transform. We only assumed that the cloak displays a symmetry of revolution about one axis in order to reduce the computational complexity of the problem. We found that while there is a strong impedance mismatch between the cloak and the anisotropic object hidden inside, an incident plane wave is nearly preserved when it interacts with this less than usual optical system (extension to completely arbitrarily shaped cloaks is a straightforward matter on the analytical side, but requires more computational power as available to us). Cloaking has been confirmed numerically for an incident plane wave in resonance with the concealed region and an analysis of the cloak's singularity has been carried out. While we focussed here on surfaces of revolution for computational convenience, the extension of the analysis to completely arbitrarily shaped cloaks with smooth boundaries is a straightforward matter. However, three-dimensional star-shaped or polyhedron cloaks would require another treatment, inspired for instance by the design of two dimensional star-shaped cloaks by blowing up a small star shaped region instead of a point as in (*17*).

### Acknowledgements

GD is thankful for a PhD scholarship from the French Ministry of Higher Education. SG acknowledges funding from the Engineering and Physical Sciences Research Council grant EPF/027125/1.